\documentclass{elsart}
\usepackage{natbib}
\begin{document}
\runauthor{Kaizoji}
\begin{frontmatter}
\title{An interacting-agent model of financial markets from the viewpoint of nonextensive statistical mechanics} 
\author{Taisei Kaizoji}
\address{Division of Social Sciences, International Christian University, Tokyo, 181-8585, Japan.}
\ead{kaizoji@icu.ac.jp}
\ead[url]{http://subsite.icu.ac.jp/people/kaizoji/}
\begin{abstract}
 In this paper we present an interacting-agent model of stock markets. We describe a stock market through an Ising-like model in order to formulate the tendency of traders getting to be influenced by the other traders' investment attitudes [1], and formulate the traders' decision-making regarding investment as the maximum entropy principle for nonextensive entropy. We demonstrate that the equilibrium probability distribution function of the traders' investment attitude is the {\it q-exponential distribution}. We also show that the power-law distribution of the volatility of price fluctuations, which is often demonstrated in  empirical studies, can be explained naturally by our model which is based on the collective crowd behavior of many interacting agents. 
\end{abstract}
\end{frontmatter}
\section{Introduction} 

 In the past decades, {\it the efficient market hypothesis} is the dominating paradigm in finance and financial engineering [1]. The efficient market hypothesis argues that the current price already contains all information and past prices can not be of help in predicting future prices. In an efficient market, stock prices are completely determined by its fundamentals, given by the present discounted value of the stream of future dividends. The prices would  be driven solely by economic news (exogenous random shocks) regarding  changes in fundamentals, so that prices follow a random walk [1]. Although numerous attempts have been made by economists to demonstrate the efficient market hypothesis, the authenticity of this hypothesis remains uncertain [2]. \footnote{We may say that the efficient market hypothesis lost ground rapidly following the accumulation of evidence against it. For instance, Shiller [3] finds that stock market volatility is far greater than could be justified by changes in dividends.}. 
A new approach proposed as an alternative to the efficient market theory is the interacting-agent approach that models the trading process with its interaction on a large ensemble of heterogeneous traders. An emerging literature on interacting-agent dynamics has been initiated by W. Brock [4], A. Kirman [5], and M. Aoki [6] and T. Lux [7]. 
A number of interacting-agent models proposed recently showed that they can generate some ubiquitous characteristics, for example the clustered volatility and the scaling behaviors of price fluctuations, found for empirical financial data as a result of interactions between agents. It seems reasonable to posit that the emergence of realistic scaling laws from based on the agents' interaction would lend convincing evidence in favor of the {\it Interacting Agent Hypothesis}. 
Although the interacting-agent models are advocated as an alternative approach to the efficient market hypothesis which is equivalent to the {\it rational expectation hypothesis} in economics [8], little attention has been given to how probabilistic rules, in which an agent switches his opinion, are connected with an agent's expectation formation. Our previous paper [9] proposed a new expectation formation hypothesis, that is, {\it the relative expectation formation hypothesis} corresponding to the interacting-agent hypothesis. The relative expectations formation of interacting agents has been formalized by using the {\it minimum average energy principle for Bolztmann-Gibbs entropy} [10, 11]. 
The aim of the present paper is to generalize the formulation of the relative expectation formation from the viewpoint of nonextensive statistical mechanics. Nonextensive statistical mechanics has been introduced as a generalization of the traditional Boltzmann-Gibbs statistical mechanics by C. Tsallis [12]\footnote{For applications of Tsallis statistics to economics, see [17-19]}. We shall present an interacting-agent model that follows the line of an Ising-like model of financial markets proposed by our previous work [9], and formulate the agents' decision-making on investment as the {\it maximum entropy principle for Tsallis entropy.} We also demonstrate that an equilibrium probability distribution of the agent's investment attitude, which is obtained as the so-called {\it q-exponential distribution}, can be derived from the relative expectations formation. We also show that the interacting-agent model provides a convincing explanation of an universal statistical characteristics of financial data, that is, the power-law tails of the distribution of volatility, measured by the absolute value of relative price changes [20]. 


\section{ An interacting-agent model} 
\subsection{The relative expectation formation}
We think of the stock market as a venue in which large numbers of traders who participate in stock trading. Traders are indexed by $ i = 1, 2, ........, N $. Traders invest their money in a stock, for example a market index traded at a price at time $ t $. They can either buy a stock or sell a stock. The traderfs investment attitude $ x_{i} $ is defined as follows: if trader $ i $ is the buyer of a stock at a time, then $ x_{i} = + 1 $, and if trader $ i $, in contrast, is the seller of a stock at a time, then $ x_{i} = - 1 $. A trader, who expects a certain exchange profit through trading, will attemp to predict every other trader's behavior in order to forecast the future movement of the price, and will choose either the same behavior or behavior contrary to behavior as the other traders' behavior. The trader's decision-making will also be influenced by changes in the news relating to the stock. When a trader hears good news about a stock, a trader may think that now is the time for him to buy the stock. . 
Formally let us assume that the investment attitude of trader $ i $ is determined by minimization of the following evaluation function $ e_i(x) $, 
\begin{equation}
    e_i(x) = - J \sum^N_{j=1} x_j x_i - b s x_i, 
   \label{eqn:a1}
\end{equation}
where $ J $ denotes the strength of the other traders' influence on the trader, and $ b $ denotes the strength of the reaction of a trader to the news $ s $ which may be interpreted as an external field, and $ x $ denotes the vector of investment attitude of all the traders $ x = (x_1, x_2,......x_N) $. A trader who has the positive value of $ J $ is called a {\it trend follower} and a trader who has the negative value of $ J $ is called a {\it contrarian}. 
The optimization problem that should be solved for every trader to achieve minimization of their evaluation functions $ e_i(x) $ is formalized by the minimization of the following function, 
\begin{equation}
   E(x) = - J \sum^N_{i=1} \sum^N_{j=1} x_i x_j 
               - \sum^N_{i=1} b s x_i. 
   \label{eqn:a2}
\end{equation}
$ E(x) $ is referred to as the internal energy in the statistical mechanics. We call the optimizing behavior of the traders interacting with the other traders a {\it relative expectation formation} [9]. For simplicity of analysis, we assume that $ b_i = 0 $. 
\subsection{A maximum entropy principle for nonextensive entropy}
In the previous subsection we assume implicitly that a trader can perceive the investment attitude of the other traders. However, in fact, a trader could not know the other traders' decisions. This is the core difficulty involved in forecasting the movement of stock markets. Under the circumstance that a large number of traders participate in trading, a probable setting may be one of best means to analyze the collective behavior of the many traders. 
Now, let us introduce a random variable $ x^k = (x^k_1, x^k_2,......, x^k_N) $, $ k = 1, 2,....., W $. The state of the traders' investment attitude $ x^k $ occurs with probability $ P_k = \mbox{Prob}(x^k) $ with the requirement $ 0 < P_k < 1 $ and $ \sum^W_{k=1} P_k = 1 $. We formulate the amount of uncertainty before the occurrence of the state $ x^k $ with probability $ P_k $ using the so-called Tsallis entropy [12] 
\begin{equation}
 S_q = - \sum^W_{i=1} P^q_k \ln_q P_k, 
\label{eqn:a3}
\end{equation}
where $ \ln_q x $ is the {\it q-logarithm function}: $ \ln_q x = (x^{1-q} - 1)/(1-q) $, $ x > 0 $. $S_q$ is a generalization of the Bolztmann-Gibbs entropy. 
Our aim is to demonstrate the equilibrium probability distribution function of the traders' investment attitude. This can be obtained by maximizing the Tsallis entropy (3) under the constraint: 
\begin{equation}
 \frac{\sum^W_{k=1} P^q_k E(x^k)}{\sum^W_{l=1} P^q_l} = U_q, \quad \sum^W_{k=1} P_k = 1. 
\label{eqn:a4}
\end{equation}
In Tsallis statistics, $ \sum^W_{k=1} P^q_k E(x^k)/(\sum^W_{l=1} P^q_l) $ is referred to as the normalized q-average energy, which is also expressed as the average energy w.r.t. the so-called escort probability 
$ P^q_k/(\sum^W_{l=1} P^q_l) $. This optimizing problem has been solved independently by Wada and Scarfone [14], and Suyari [15]. They derived an alternative expression, which is non-self-referential, for the Tsallis distribution [13]. 
The first-order conditions for a minimum are derived by setting the partials of the Lagrangian 
\begin{equation}
  \phi_q(P_k, \alpha, \beta) = S_q - \alpha (\sum^W_{k=1} P_k - 1) - \beta \frac{\sum^W_{k=1} P^q_k (E(x^k) - U_q)}{\sum^W_{l=1} P^q_{l}} 
\label{eqn:a5}
\end{equation}
with respect to $ P_k $, $ \alpha $ and $ \beta $ equal to zero: 
\begin{equation}
 \frac{\partial \phi_q}{\partial P_k} = - qP^{q-1}_{k} \ln_{q} P_k - 1 - \alpha - \beta \frac{qP^{q-1}_k (E(x^k) - U_q)}
{\sum^W_{l=1} P^q_l}  = 0
\label{eqn:a6}
\end{equation}
\begin{equation}
 \frac{\partial \phi}{\partial \alpha} =  1 - \sum^W_{k=1} P_k = 0 
\label{eqn:a7}
\end{equation}
\begin{equation}
 \frac{\partial \phi}{\partial \beta} = - \frac{\sum^W_{k=1} P^q_k (E(x^k) - U_q)}{\sum^W_{l=1} P^q_{l}} = 0. 
\label{eqn:a8}
\end{equation}
The equilibrium probability distribution function as the solution of the above optimization problem is 
\begin{equation}
 P_k = \frac{1}{Z_q(\beta_q)} \exp_{q}(-\beta_q (E(x^k) - U_q))
\label{eqn:a9}
\end{equation}
where $ Z_q(\beta_q) $ is the generalized partition function, 
\begin{equation}
Z_q(\beta_q) = \sum^W_{k=1} \exp_q(- \beta_q (E(x^k) - U_q)).
\label{eqn:a10}
\end{equation} 
$ \beta_q $ is defined by 
$$ \beta_q = \frac{q}{q + (1 + \alpha) (1 - q)} \beta. $$
$ \exp_q(x) $ is the {\it q-exponential function} defined by 
$$ \exp_q(x) = \left\{\begin{array}{rl} 
                 [1 + (1 - q) x]^{1/(1-q)} & \quad 
                 \mbox{if $ 1 + (1 - q) x > 0 $}. \\ 
                 0   &  \quad \mbox{otherwise}
                 \end{array}\right. $$ 
The equilibrium probability distribution function $ P_k $ is called the {\it q-exponential distribution}, where $ P_k $ is the probability that the traders' investment attitude is in the state $ k $. The detailed derivation of (9) is given by Suyari [16]. 
As is well known, for $ q > 1 $ the q-exponential distribution (9) for a large $ E(x^k) $ value is approximated by a power-law distribution,
\begin{equation}
 P_k \propto E(x^k)^{1/(1 - q)}.
\label{eqn:a11}
\end{equation}
\subsection{The price adjustment processes}
Generally, the price changes in financial market are subject to the law of supply and demand; i.e., the price rises when there is excess demand, and the price falls when there is excess supply. Since traders are supposed to either buy or sell a stock at a ptime, the price will rise if the number of buyers exceeds the number of sellers ($ \sum^N_{i=1} x_i > 0 $) because there may be excess demand, and the price will fall if the number of seller exceeds the number of buyers ($ \sum^N_{i=1} x_i < 0 $) because there may be excess supply. 
It is natural to consider that the price fluctuations measured by relative price change are proportional to the demand-supply gap, so that {\it volatility}, measured by the absolute value of relative price changes, is proportional to the absolute value of the demand-supply gap. 
Since the difference between the number of buyers and the number of sellers is proportional to the value of $ \sum^N_{i=1} \sum^N_{j=1} x_i x_j $, the absolute value of the demand-supply gap is proportional to the value of $ \sum^N_{i=1} \sum^N_{j=1} x_i x_j $. Volatility is then described as 
\begin{equation}
V(t+1) = |\ln p(t+1) - \ln p(t)|  \propto \sum^N_{i=1} \sum^N_{j=1} x_i(t) x_j(t) = - \frac{1}{J} E(x(t)). 
\label{eqn:a11}
\end{equation} 
where $ p(t) $ denotes the price at time $ t $. 
\subsection{Volatility distribution} 
Let us assume that the traders are the {\it contrarians}, that is, $ J < 0 $. For large values of volatility, the volatility distribution is approximated by a power-law distribution, 
\begin{equation}
 P(V) \propto V^{1/(1 - q)}, 
\label{eqn:a13}
\end{equation}
where $ V $ denotes volatility defined by (11). Empirical studies on volatility in stock markets often determine the power-law distributions of volatility. Liu, et. al. [20] analyzed the SP 500 stock index and the stock prices of the individual companies comprising the SP 500 for the 13-year period from January 1984 to December 1996, and found that the distribution of the volatility is consistent with asymptotic power-law behavior, characterized by an exponent $4$. Therefore, the parameter $ q $ of the volatility distribution (13) is estimated as $ q = 1.25 $ for the U.S. stock market. 

\section{Concluding remarks}
This paper presents an interacting-agent model of speculative activity explaining the power laws of the volatility distribution in terms of nonextensive statistical mechanics. We show theoretically that the power-laws governing price fluctuations can be easily understood as a result of the relative expectations formation of many interacting agents, formalized as the maximum entropy principle for nonextensive entropy. 
\section{Acknowledgement} 
I wish to thank Prof. Hiroki Suyari for valuable advice. All remaining errors, of course, are mine. This research was supported in part by a grant from the Japan Society for the Promotion of Science under the Grant-in-Aid, No. 06632.

\end{document}